\documentclass[epj]{webofc}
\usepackage[varg]{txfonts}   
\usepackage{graphicx,color} 
\usepackage{bm}       
\usepackage{amsmath}  
\usepackage{amssymb}

\newcommand{\im}[1]{\mathrm{Im}\left(#1\right)}
\newcommand{\re}[1]{\mathrm{Re}\left(#1\right)}

\newcommand{\omb}{\omega_\mathrm{B}}
\newcommand{\om}{\omega}

\newcommand{\abar}{\bar a}
\newcommand{\Vbar}{\bar{V}}
\newcommand{\Vtil}{\widetilde{V}}
\newcommand{\Bbar}{\bar{B}}
\newcommand{\Btil}{\widetilde{B}}
\newcommand{\btil}{\tilde{b}}

\woctitle{21st International Conference on Few-Body Problems in Physics}
\begin{document}
\title{Induced two-body scattering resonances from a \\square-well potential with oscillating depth}
\author{D.~Hudson Smith\inst{1}\fnsep\thanks{\email{smith.7991@osu.edu}}}

\institute{Department of Physics,
         The Ohio State University, Columbus, OH\ 43210, USA\\}

\abstract{
In systems of ultracold atoms, pairwise interactions can be resonantly enhanced by a new mechanism which does not rely upon a magnetic Feshbach resonance. In this mechanism, interactions are controlled by tuning the frequency of an oscillating parallel component of the magnetic field close to the Bohr frequency for the transition to a two-atom bound state. The real part of the s-wave scattering length $a$ has a resonance as a function of the oscillation frequency near the Bohr frequency. The resonance parameters can be controlled by varying the amplitude of the oscillating field. The amplitude also controls the imaginary part of $a$ which arises predominantly because the oscillating field converts atom pairs into molecules. For the case of a shallow bound state in the scattering channel, the dimensionless resonance parameters are universal functions of the dimensionless oscillation amplitude.
}

\maketitle
\section{Introduction}\label{sec:intro}
A unique feature of ultracold atomic physics is the ability to precisely control the interactions among particles all the way from zero interactions to infinitely attractive or repulsive interactions. This tunability has led to many breakthroughs in few- and many-body physics. In most current experiments, interactions are controlled by exploiting a {\it magnetic Feshbach resonance} (MFR), where an external magnetic field is tuned near the value $B_0$ where a pair of unbound atoms becomes degenerate with a two-atom bound state \cite{Review_FR}. For ultracold atoms, the strength of interactions is determined by the s-wave scattering length $a$. Near $B_0$, $a$ is a simple function of the magnetic field $B$:
\begin{equation}\label{eq:feshbach}
\frac{1}{a(B)}=\frac{1}{a_{\mathrm{bg}}}\,\frac{B-B_0}{B-B_0-\Delta}+i\gamma,
\end{equation}
where $a_{\mathrm{bg}}$ is the background scattering length, $\Delta$ is the width of the resonance, and $\gamma$ is non-zero only if the colliding atoms have a spin-relaxation scattering channel. 

In this letter, we examine a new mechanism, {\it modulated-magnetic Feshbach resonance} (MMFR), for resonantly enhancing the scattering length in ultracold gases. This mechanism is related to {\it modulated-magnetic spectroscopy} or {\it wiggle spectroscopy} which has been used to measure molecular binding energies and other properties for several alkali-metal atoms \cite{Wieman2005,Inguscio0808,Chin_2009,Khaykovich1201,Hulet1302}. A MMFR is induced by an oscillating magnetic field, $B(t)=\Bbar + \Btil\cos(\om t)$, that is parallel to the spin-quantization axis of the atoms and near the Bohr frequency for the transition between an atom pair and a bound state. Near a MFR, the Bohr frequency can be decreased by adjusting $\Bbar$ to bring the resonance molecule closer to the threshold. Previous theoretical treatments involving oscillating magnetic fields have focused on molecule formation \cite{Hanna:pra:2007,Langmack:prl:2015,Mohapatra:pra:2015}. In constrast, the present work focuses on the effect of the oscillating magnetic field on the elastic scattering properties of the atoms.

For $\omega$ near the Bohr frequency $\omb$, the scattering length is a simple function of $\omega$:
\begin{equation}\label{eq:resonance_eqn}
\frac{1}{a(\om)}=\frac{1}{\abar}\,\frac{\om-\om_0}{\om-\om_0-\delta}+i\gamma,
\end{equation}
where $\abar$ is the scattering length in the absence of the modulated field, $\omega_0$ is close to $\omb$, and $\delta$ is the width of the resonance. The inverse scattering length has been given a frequency-independent positive imaginary part $\gamma$ that is only important very close to the resonance. The parametrization in Eq.~\eqref{eq:resonance_eqn} ensures that $\im{a}\leq 0$ for all $\omega$, as required by unitarity. The maximum value of $|\re{a}|$ is $1/2\gamma$. The imaginary part of $a$ arises from collisions in which a pair of low-energy atoms emits or absorbs one or more quanta of frequency $\omega$ and forms a molecule or an excited scattering state. Near resonance, the imaginary part is dominated by molecule formation.

We calculate $a(\omega)$ for a short-range potential with an oscillating depth. We show the existence of a resonance near $\omega=\omb$, and we extract the dependencies of the resonance parameters $\delta$, $\Delta\om_0=\om_0-\omb$, and $\gamma$ on the amplitude of the oscillating part of the potential. For small amplitudes, the resonance parameters each scale as the square of the amplitude. We examine the special case where the resonance molecule is a shallow dimer in the scattering channel. In this case, the dimensionless resonance parameters $\delta/\omb$, $\Delta\om_0/\omb$, and $\gamma\abar$ are universal functions of a dimensionless oscillatory magnetic field variable that is proportional to $\Btil$.

\section{Single-channel scattering model}\label{sec:model}
We consider a single-channel model for the effects of the oscillating magnetic field.  For atomic separations $r$ greater than $r_0$, the atoms are non-interacting. For $r<r_0$, the atoms interact through a square-well potential with an oscillating depth. The s-wave scattering properties for atoms with mass $m$ are extracted from the solution to the time-dependent Schr{\"o}dinger equation for $u(r,t)=rR(r,t)$, where $R(r,t)$ is the radial wave-function for the relative position of the atom pair with an interaction potential $(\Vbar+\widetilde{V}\cos(\omega t))\theta(r_0-r)$. The time-averaged depth of the potential, $\Vbar$, and the range, $r_0$, are chosen such that the time-averaged potential supports a bound state with Bohr frequency $\omb$. The term proportional to $\Vtil$ represents the time-dependent shift in the binding energy of the resonance molecule relative to the scattering threshold. For a shallow bound state in the scattering channel, this energy shift is the change in the binding energy of the molecule when the magnetic field is shifted by $\Btil$. For a bound state in another hyperfine channel, the energy shift is $-\delta\mu\,\Btil\cos(\om t)$, where $\delta\mu$ is the difference between the magnetic moment of the bound state and the magnetic moment of two atoms.

The Schr{\"o}dinger equation can be solved by using Floquet's theorem as in Ref.~\cite{Reichl:1999prb}. This theorem asserts that for time-periodic potentials the wavefunction can be rewritten
$u(r,t)=e^{-iE_\mathrm{F}t}\phi(r,t)$,
where the Floquet eigenvalue $E_\mathrm{F}$ is determined by the boundary conditions, and $\phi(r,t)$ has the same periodicity as the time-dependent part of the potential. The wavefunction for $r>r_0$ contains an incoming mode with energy $k^2/m$, where $k$ is the small relative momentum of the atom pair. The requirement that $u(r,t)$ be continuous at $r=r_0$ implies that $E_\mathrm{F}=k^2/m+j\pi\omega$ where $j$ is any integer (taken to be zero below for simplicity). The Schr{\"o}dinger equation can be solved analytically giving
\begin{equation}\label{eq:usol} 
 u(r,t) = \sum\limits_{n=-\infty}^{\infty} \\
  \begin{cases}
   2ia_n \sin(q_nr)\exp\left[{-i(k_n^2/m)t+i\Vtil\sin(\omega t)/\omega}\right]
    & r < r_0, \\
   (A^\mathrm{out}_n e^{ik_n r}+A^\mathrm{in}_n e^{-ik_n r})\exp\left[{-i(k_n^2/m)t}\right]
   & r \geq r_0,
  \end{cases}
\end{equation}
where $k_n=[k^2+mn\omega]^{1/2}$ and $q_n=[k^2+m(\Vbar+n\omega)]^{1/2}$. The solution for $r>r_0$ is a superposition of freely propagating modes. The coefficients $A^\mathrm{out}_n$ and $A^\mathrm{in}_n$ represent the amplitudes of outgoing and incoming modes, respectively. The $n=0$ mode corresponds to the low-energy scattering state with energy $k^2/m$. The additional so-called {\it Floquet modes} with energies that differ by integer multiples of $\omega$ are necessary to satisfy the boundary conditions at $r=r_0$. Physically speaking, incident particles scattering with energy $k^2/m$ can absorb (emit) quanta from (to) the oscillating field. They will, therefore, emerge as a superposition of modes with energies that differ from $k^2/m$ by integer multiples of $\omega$. The negative energy modes are exponentially damped for $r>r_0$, and they do not propagate. Both $u(r,t)$ and $\partial u(r,t)/\partial r$ must be continuous at $r=r_0$. These boundary conditions can be used to eliminate the $a_n$ coefficients and find the $S$-matrix which relates the amplitudes of the incoming and outgoing modes. The scattering length is then extracted from the $S$-matrix element corresponding to low-energy elastic scattering (For a more detailed derivation, see Ref.~\cite{Smith:arxiv:2015})

\begin{figure}
\centering
\begin{minipage}{0.45\textwidth}
	\includegraphics[width=\columnwidth]{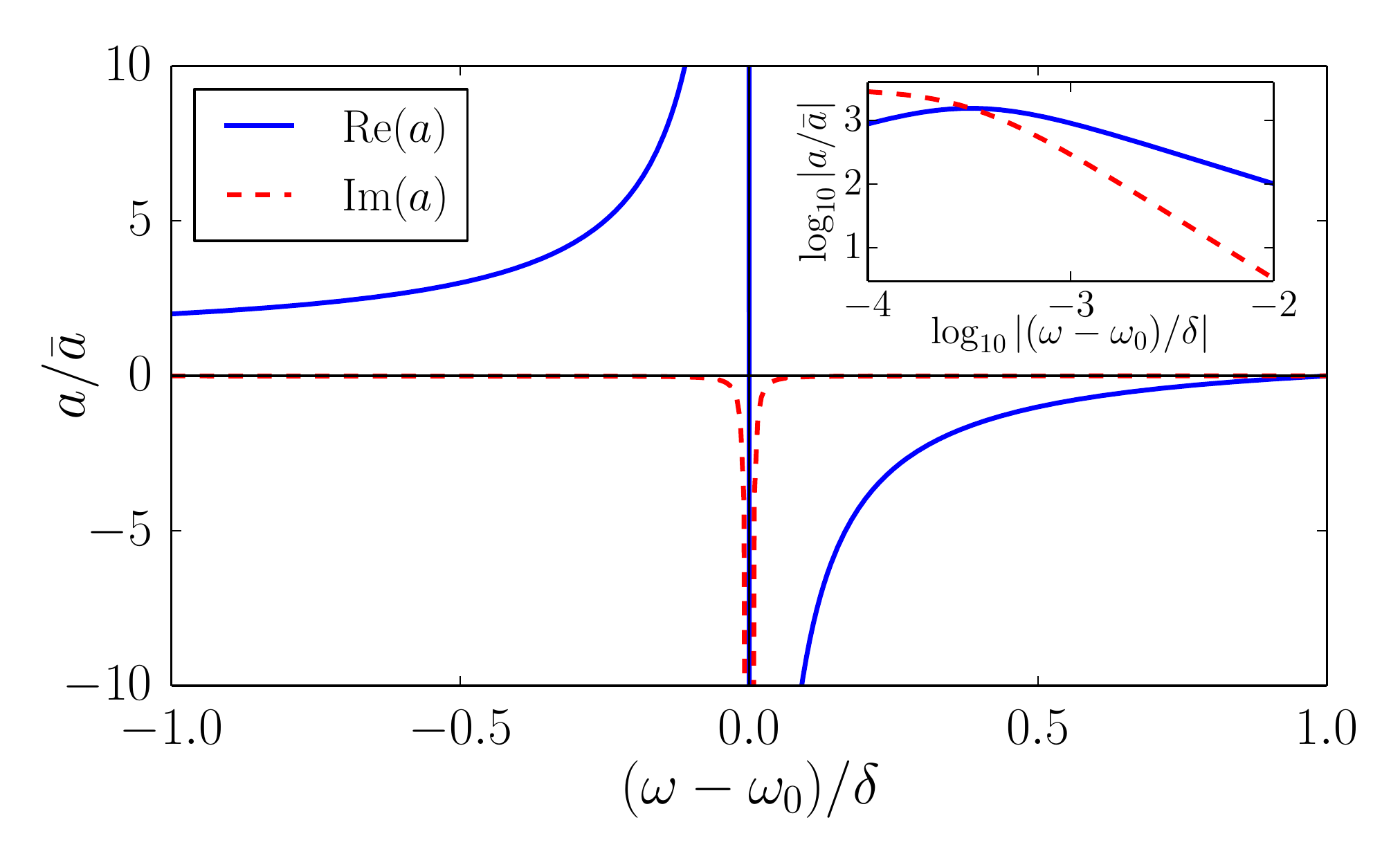}\vspace{-15px}
	\caption{The real and imaginary parts of $a$ as functions of $\omega$ for a MMFR where the resonance molecule is a shallow dimer in the scattering channel. The value of $\btil$ is $0.05$. The inset shows the absolute values of the same quantities on a logarithmic scale.}
	\label{fig:scatLengthZoomed}
\end{minipage}\hfill
\begin{minipage}{0.45\textwidth}
	\includegraphics[width=\columnwidth]{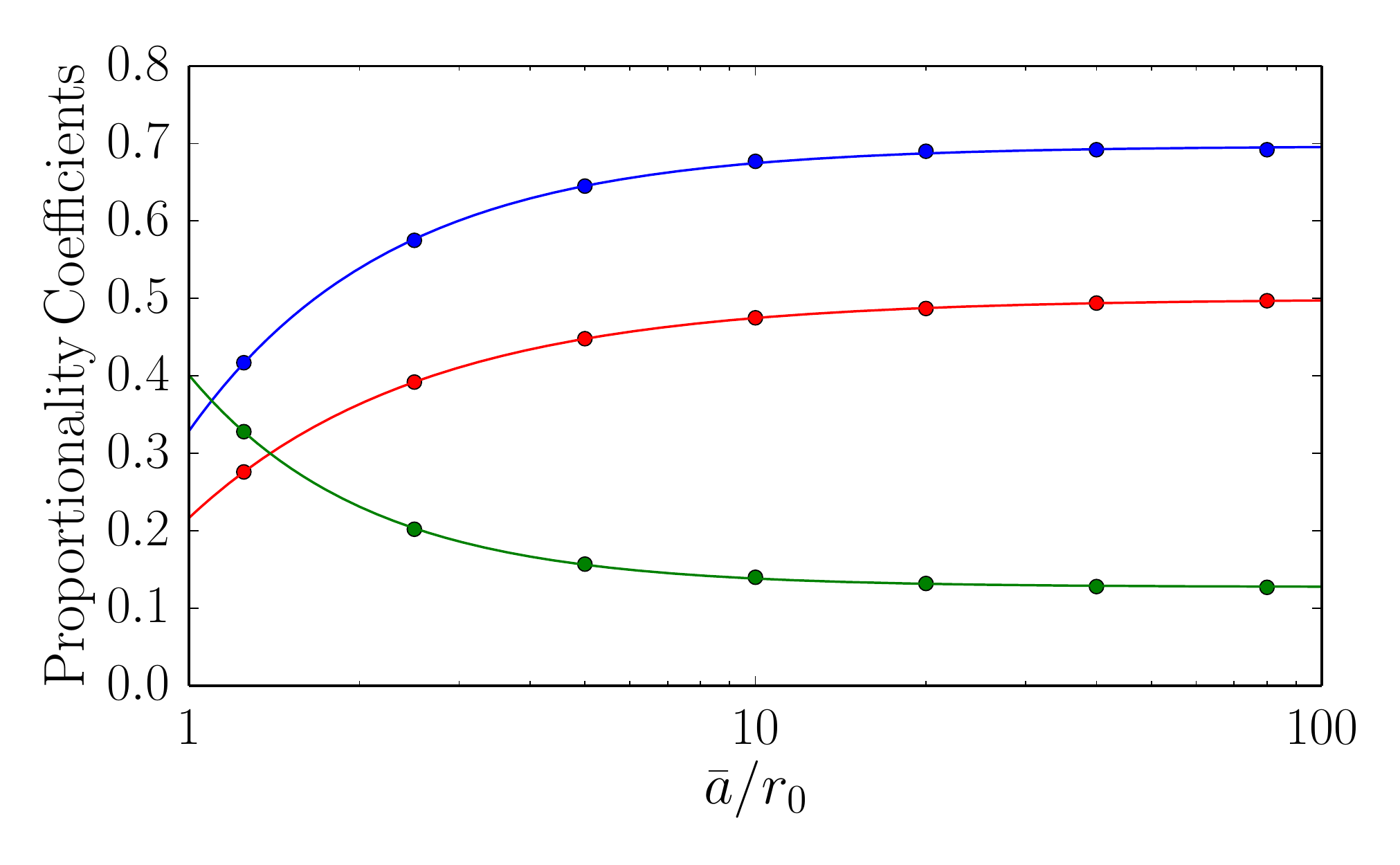}\vspace{-15px}
	\caption{Numerical results for the coefficients of $\btil^2$ for the dimensionless resonance parameters $\Delta\om_0/\omb$, $\delta/\omb$, and $\gamma\abar$ (top to bottom) as functions of $\abar$ along with power-law fits to guide the eye. In the large $\abar$ limit, the coefficients approach the universal numbers given in Eqs.~\eqref{eq:scaling_behavior}.}
	\label{fig:universallimit}
\end{minipage}
\vspace{-20px}
\end{figure}

Figure \ref{fig:scatLengthZoomed} plots the real and imaginary parts of $a$ as functions of frequency, demonstrating the familiar Feshbach resonance shape. Molecule formation limits the maximum value of the real part of $a$. For small $\Vtil$ the resonance parameters $\delta$, $\Delta\om_0$, and $\gamma$ are proportional to $\Vtil^2$ with coefficients that depend upon the parameters of the potential.

\section{Shallow dimer in the scattering channel}\label{sec:dimer}
The single-channel model is well-suited for describing an MMFR for which the resonance molecule is a shallow dimer in the scattering channel. The dimer's binding energy is $\omb=1/m\abar^2$ where $\abar=a(\Bbar)$ is given by Eq.~\eqref{eq:feshbach} with $\gamma=0$. We choose $\Vbar$ such that the binding energy of the shallowest molecule $E(\Vbar)$ is $\omb$. We choose $\Vtil$ such that the change in $E(\Vbar)$ from a small modification $\Vtil$ to the depth of the potential equals the change in $\omb$ from a small modification $\Btil$ to the amplitude of the magnetic field. Equating these energy shifts, 
$E'(\Vbar)\,\Vtil=-2\omb\,\btil$,
where we have introduced the dimensionless oscillatory magnetic field variable $\btil = [a'(\Bbar)/\abar]\Btil$. Inserting the value of $\Vtil$ determined by equating the energy shifts into Eq.~\eqref{eq:usol}, we find the change in the scattering length resulting from a small oscillation of the magnetic field.

We now extract the dependencies of the resonance parameters on $\btil$ by fitting the numerical results for $a$ extracted from Eq.~\eqref{eq:usol} with the analytic parametrization in Eq.~\eqref{eq:resonance_eqn} for different values of $\btil$. The resonance parameter $\abar$ is set equal to $a(\Bbar)$. The parameters $\delta$ and $\omega_0$ are determined by fitting the real part of the inverse scattering length to the real part of Eq.~\eqref{eq:resonance_eqn}. The parameter $\gamma$ is then determined by calculating $a$ at $\om=\om_0$. For small $\btil$, the dimensionless resonance parameters $\delta/\omb$, $\Delta\om_0/\omb$, and $\gamma \abar$ are proportional to $\btil^2$. In the limit $\abar \gg r_0$ the proportionality coefficients are universal numbers that are independent of the short-range potential:
\begin{align}\label{eq:scaling_behavior}
\delta/\omb &= 0.50\,\btil^2, \nonumber \\
\Delta\om_0/\omb &= 0.69\,\btil^2, \nonumber \\
\gamma\abar &= 0.13\,\btil^2.
\end{align}
Figure \ref{fig:universallimit} plots the proportionality coefficients as functions of $\abar$. From $\im{a}$, we can predict the on-resonance dimer formation rate. This rate agrees with the result from Fermi's Golden Rule \cite{Langmack:prl:2015}.

\section{Conclusion}\label{sec:summary}
We have examined a new mechanism for resonantly enhancing the s-wave scattering length $a$ by tuning the frequency of an oscillating magnetic field that is parallel to the spin-quantization axis. Along with enhancing the real part of $a$, the oscillating field also generates an imaginary part associated with molecule formation. The real and imaginary parts of $a$ can be controlled by the frequency and amplitude of the oscillating magnetic field. A simple model was used to determine how the resonance parameters scale with the amplitude of the oscillating field. For the special case where the resonance molecule is a shallow dimer in the scattering channel, the dimensionless resonance parameters are universal functions of $\btil$.

\begin{acknowledgement}
This research was supported by the National Science Foundation under grant PHY-1310862. Many thanks are due to Dr.~Cheng Chin whose insightful question sparked this research topic and to Logan Clark for helpful discussions. I also thank Dr.~Eric Braaten for invaluable feedback during the writing of this paper.
\end{acknowledgement}

%
%
%

\end{document}